\documentclass[a4paper,11pt]{article}
\pdfoutput=1 

\usepackage{jcappub} 

\usepackage[T1]{fontenc}
\usepackage[utf8]{inputenc}
\usepackage{caption}
\usepackage{subcaption}

\usepackage{diagbox}
\usepackage{color}
\usepackage{multirow}
\usepackage{hhline}
\usepackage{tabularx}

\title{Space-borne atom interferometric gravitational wave detections. Part II. Dark sirens and finding the one}

\author[a]{Tao Yang}
\emailAdd{yangtao.lighink@gmail.com}

\author[a]{Hyung Mok Lee}
\emailAdd{hmlee@snu.ac.kr}

\author[b,c,d]{Rong-Gen Cai}
\emailAdd{cairg@itp.ac.cn}

\author[a]{Han Gil Choi}
\emailAdd{alivespace@snu.ac.kr}

\author[e,f]{Sunghoon Jung}
\emailAdd{sunghoonj@snu.ac.kr}

\affiliation[a]{Astronomy Program Department of Physics and Astronomy, Seoul National University, 1 Gwanak-ro, Gwanak-gu, Seoul 08826, Korea}

\affiliation[b]{CAS Key Laboratory of Theoretical Physics, Institute of Theoretical Physics, Chinese Academy of Sciences, Beijing 100190, China}
\affiliation[c]{School of Physical Sciences, University of Chinese Academy of Sciences, No.19A Yuquan Road, Beijing 100049, China}
\affiliation[d]{School of Fundamental Physics and Mathematical Sciences, Hangzhou Institute for Advanced Study (HIAS), University of Chinese Academy of Sciences, Hangzhou 310024, China}

\affiliation[e]{Center for Theoretical Physics, Department of Physics and Astronomy, Seoul National University, Seoul 08826, Korea}
\affiliation[f]{Astronomy Research Center, Seoul National University, Seoul 08826, Korea}

\abstract{
In this paper, we investigate the potential of dark sirens by the space-borne atom interferometric gravitational-wave detectors to probe the Hubble constant. In the mid-frequency band, the sources live a long time. The motion of a detector around the Sun as well as in Earth orbit would induce large Doppler and reorientation effects, providing a precise angular resolution. Such precise localization for the GW sources makes it possible to observe the dark sirens with only one potential host galaxy, which are dubbed ``golden dark sirens''.  We construct the catalogs of golden dark sirens and estimate that there are around 79 and 35 golden dark sirens of binary neutron stars (BNS) and binary black holes (BBH) that would be pass the detection threshold of AEDGE in 5 years. Our results show that with 5, 10, and all 79 golden dark BNS tracked by AEDGE one can constrain $H_0$ at 5.5\%, 4.1\%, and 1.8\% precision levels. With 5, 10, and all 35 golden dark BBH one can constrain $H_0$ at 2.2\%, 1.8\%, and 1.5\% precision levels, respectively. It suggests that only 5-10 golden dark BBH by AEDGE are sufficient to arbitrate the current tension between local and high-$z$ measurements of $H_0$.
}

\keywords{gravitational waves / theory, gravitational wave detectors, gravitational waves / experiments, dark energy theory}

\begin{document}
\maketitle
\flushbottom

\section{Introduction}

Atom interferometers (AIs) as the novel gravitational wave (GW) detectors have been proposed a decade ago~\cite{Dimopoulos:2007cj,Dimopoulos:2008sv,Graham:2012sy,Hogan:2015xla}. Unlike the traditional laser interferometers (LIs) such as LIGO~\cite{LIGOScientific:2007fwp,aLIGO:2016pgl}, in the concept AIs, gravitational radiation is sensed through precise measurement of the light flight time between two distantly separated atomic inertial references, each in a satellite in Medium Earth orbit (MEO). Ensembles of ultra-cold atomic Sr atoms at each location serve as precise atomic clocks. Light flight time is measured by comparing the phase of laser beams propagating between the two satellites with the phase of lasers referenced to the Sr optical transitions~\cite{Graham:2017pmn}. The AI detector projects such as ground based ZAIGA~\cite{Zhan:2019quq} in China, AION~\cite{Badurina:2019hst} in the UK, MIGA~\cite{Geiger:2015tma} in France, ELGAR~\cite{Canuel:2019abg} in Europe, and the space-borne MAGIS~\cite{Graham:2017pmn} and AEDGE~\cite{AEDGE:2019nxb} have been proposed and in preparation. 

The success of LIs detector LIGO on discovering the first gravitational wave event marking the new era of GW multi-messenger astronomy~\cite{LIGOScientific:2016aoc}. Up to now, LIGO has reported more than 50 confirmed GW events produced by the merger of the binary black holes (BBH), of the binary neutron stars (BNS), and of the neutron star-black hole binary (NS-BH)~\cite{LIGOScientific:2016aoc,LIGOScientific:2017vwq,LIGOScientific:2018mvr,LIGOScientific:2020ibl,LIGOScientific:2021qlt}. GWs  play significant roles in  modern cosmology, astrophysics and fundamental physics (see reviews~\cite{Schutz:1999xj,Barack:2018yly,Sasaki:2018dmp,Gair:2012nm,Ezquiaga:2018btd,Cai:2017cbj,Meszaros:2019xej,Christensen:2018iqi,Perkins:2020tra}, and a very recent review of the progress in GW physics~\cite{Bian:2021ini}). One of the most important applications of GWs on cosmology is measuring the cosmological parameters like Hubble constant~\cite{Schutz:1986gp}. This property dubbed ``standard sirens'' makes gravitational wave one of the most promising probes to resolve the Hubble tension which arises from the discrepancy of the measurement of $H_0$ by Cosmic Microwave Background (CMB) observations and Type Ia supernovae (SNe Ia) via the local distance ladder. Currently this tension is greater than 4$\sigma$ confidence level (for the summary of $H_0$ measurements in different manners and the possible solutions for this tension see reviews~\cite{Verde:2019ivm,DiValentino:2021izs}). The third-party measurement of $H_0$ independent of CMB and local distance ladder is thus urgent and crucial. The observation of gravitational wave (GW170817) from a BNS with its electromagnetic (EM) counterparts made the first attempt to measure the Hubble constant from GW standard siren~\cite{LIGOScientific:2017adf}. Though currently, it is not precise enough to resolve the Hubble tension, the forecasting for future LIGO network shows that a 2 percent Hubble constant measurement can be obtained within 5 years, which is sufficient to arbitrate the Hubble tension~\cite{Chen:2017rfc} (several examples of forecasting the applications of standard sirens on cosmology can also be found in~\cite{Cai:2016sby,Cai:2017yww,Yang:2021qge}). 

AIs, as novel GW detectors, can probe both dark matter and gravitational waves (see the introductions in~\cite{Graham:2017pmn,Badurina:2019hst,AEDGE:2019nxb}). They are the candidates to probe GW in the Deci-Hz gap between LIGO/Virgo and LISA. In our first paper for applications of AIs on cosmology~\cite{Cai:2021ooo} (hereafter Paper I), we focus on the bright sirens, i.e., the joint observations of GWs with their EM counterparts (though the EM counterparts are identified by the follow-up observations). We forecast the ability of bright siren by the space-borne AEDGE on constraining cosmological parameters. In the bright siren case, one usually considers BNS (or NS-BH binaries) with their associated short gamma-ray burst (GRBs) to identify the host galaxy and hence the redshift information of the source~\cite{Dalal:2006qt,Nissanke:2009kt,Zhao:2010sz}. The application of GW170817 and its EM counterparts on measuring Hubble constant~\cite{LIGOScientific:2017adf} has proved this feasibility. While in the dark siren case (the GWs without EM counterparts), one can only adopt a statistical way to infer the redshift of the host galaxy. Several statistical approaches have been proposed and the applications on cosmology have been investigated~\cite{DelPozzo:2011vcw,Taylor:2011fs,Chen:2016tys,Chen:2017rfc,Yu:2020vyy,Borhanian:2020vyr,Mukherjee:2020hyn,Wang:2020dkc}. The measurements of the Hubble constant from current dark sirens have also been reported~\cite{LIGOScientific:2018gmd,DES:2019ccw,LIGOScientific:2019zcs,DES:2020nay}. Though the constraint of the Hubble constant by a single dark siren is much looser than that by a bright siren, the huge number of dark sirens can compensate for this inferiority. In this paper, we extend our study of AIs to the dark siren case. For AIs, only a single baseline is required to sense gravitational waves. However, in the mid-frequency band, the time scales of the inspiral phase of BNS and BBH are from months to years. The measurement baseline reorients on a rapid time scale compared to the observation duration. As a detector reorients and/or moves, the observed waveform and phase are modulated and Doppler-shifted. This allows efficient determination of sky position and polarization information~\cite{Graham:2017lmg,Graham:2017pmn}. The detailed study of the localization of GW sources by AIs can be found in~\cite{Graham:2017lmg}. The very precise localization of the GW sources inspires us to investigate the potential of AIs' dark sirens on cosmology, especially for the measurement of the Hubble constant.

We adopt a similar method as in Paper I to construct the catalogs of simulated GWs. We consider both BNS and BBH which are not associated with EM counterparts. In the mid-frequency band between 0.01 and a few Hz, AIs can also observe GWs produced by the intermediate-mass black hole (IMBH) and NS-BH binaries. However, since we do not have enough informative knowledge of the properties (mass range, merger rates, etc.) of IMBH and NS-BH binaries~\cite{LIGOScientific:2021tfm,LIGOScientific:2021qlt}, it is hard to simulate their catalogs. We will see later that in the resonant modes, AIs can only track a small fraction of the total GWs that pass the detection threshold of AEDGE. These factors make us only focus on the BNS and BBH cases and neglect the contributions from IMBH and NS-BH binaries whose merger rates are relatively less.

In this paper, we continue using AEDGE~\cite{AEDGE:2019nxb} as our fiducial space-borne AIs whose mission duration is from 5 years to 10 years. In the mid-frequency band, the motion of the detectors around the Sun as well as in Earth orbit would provide a very precise angular localization by tracking the long-time inspiral phase of BNS and BBH~\cite{Graham:2017lmg}. Such precise localization for the GW sources makes it possible to find the ``golden dark sirens'', i.e., the GWs have only one potential host galaxy in their sky localization volumes~\cite{Chen:2016tys}. Since the number of dark sirens that can be tracked by AEDGE in the resonant modes is very limited, in this paper, we focus on the golden dark sirens to measure the Hubble constant. 

The structure of this paper is as follows. In section~\ref{sec:catalog} we construct the catalogs of BNS and BBH. For each event, we calculate its localization volume. By assuming the number density of the galaxies, we select the golden dark BNS and BBH in the catalogs. We then construct the Hubble diagram of the golden dark sirens. In section~\ref{sec:hubble} we forecast the measurements of Hubble constant by assuming different numbers of golden dark BNS and BBH that AEDGE can actually track. We give our conclusions and discussions in section~\ref{sec:conclusion}.

\section{The catalogs of BNS and BBH dark sirens \label{sec:catalog}}
To sample the catalogs of BNS and BBH dark sirens we first need to draw the redshift distribution of such binaries. In this paper, we assume the formation of compact binaries tracks the star formation rate.
The merge rate per volume at a specific redshift $R_m(z_m)$ is related to the formation rate of massive binaries and the time delay distribution $P(t_d,\tau)=\frac{1}{\tau}\exp(-t_d/\tau)$ with an e-fold time of $\tau=100$ Myr~\cite{Vitale:2018yhm},
\begin{equation}
R_m(z_m)=\int_{z_m}^{\infty}dz_f\frac{dt_f}{dz_f}R_f(z_f)P(t_d) \,.
\label{eq:Rm}
\end{equation}
Here $t_m$ (or the corresponding redshift $z_m$) and $t_f$ are the look-back time when the systems merged and formed. $t_d=t_f-t_m$ is the time delay. $R_f$ is the formation rate of massive binaries and we assume it is proportional to the Madau-Dickinson (MD) star formation rate~\cite{Madau:2014bja},
\begin{equation}
\psi_{\rm MD}=\psi_0\frac{(1+z)^{\alpha}}{1+[(1+z)/C]^{\beta}} \,,
\label{eq:psiMD}
\end{equation}
with parameters $\alpha=2.7$, $\beta=5.6$ and $C=2.9$. The coefficient $\psi_0$ is the normalization factor which is determined by the merger rate at $z=0$. We adopt $R_m(z=0)$ of BNS and BBH as the local merger rates inferred from GWTC-2~\cite{LIGOScientific:2020ibl,LIGOScientific:2020kqk}, which have been updated by including the O3 first half run. The local merger rates of BNS and BBH are $\mathcal{R}_{\rm BNS}=320^{+490}_{-240}~\rm Gpc^{-3}~\rm yr^{-1}$ and $\mathcal{R}_{\rm BBH}=23.9^{+14.3}_{-8.6}~\rm Gpc^{-3}~\rm yr^{-1}$~\cite{LIGOScientific:2020kqk}. Then we convert the merger rate per volume in the source frame to merger rate density per unit redshift in the observer frame
\begin{equation}
R_z(z)=\frac{R_m(z)}{1+z}\frac{dV(z)}{dz} \,,
\label{eq:Rz}
\end{equation}
where $dV/dz$ is the comoving volume element. 

\begin{figure}
\centering
\includegraphics[width=0.49\textwidth]{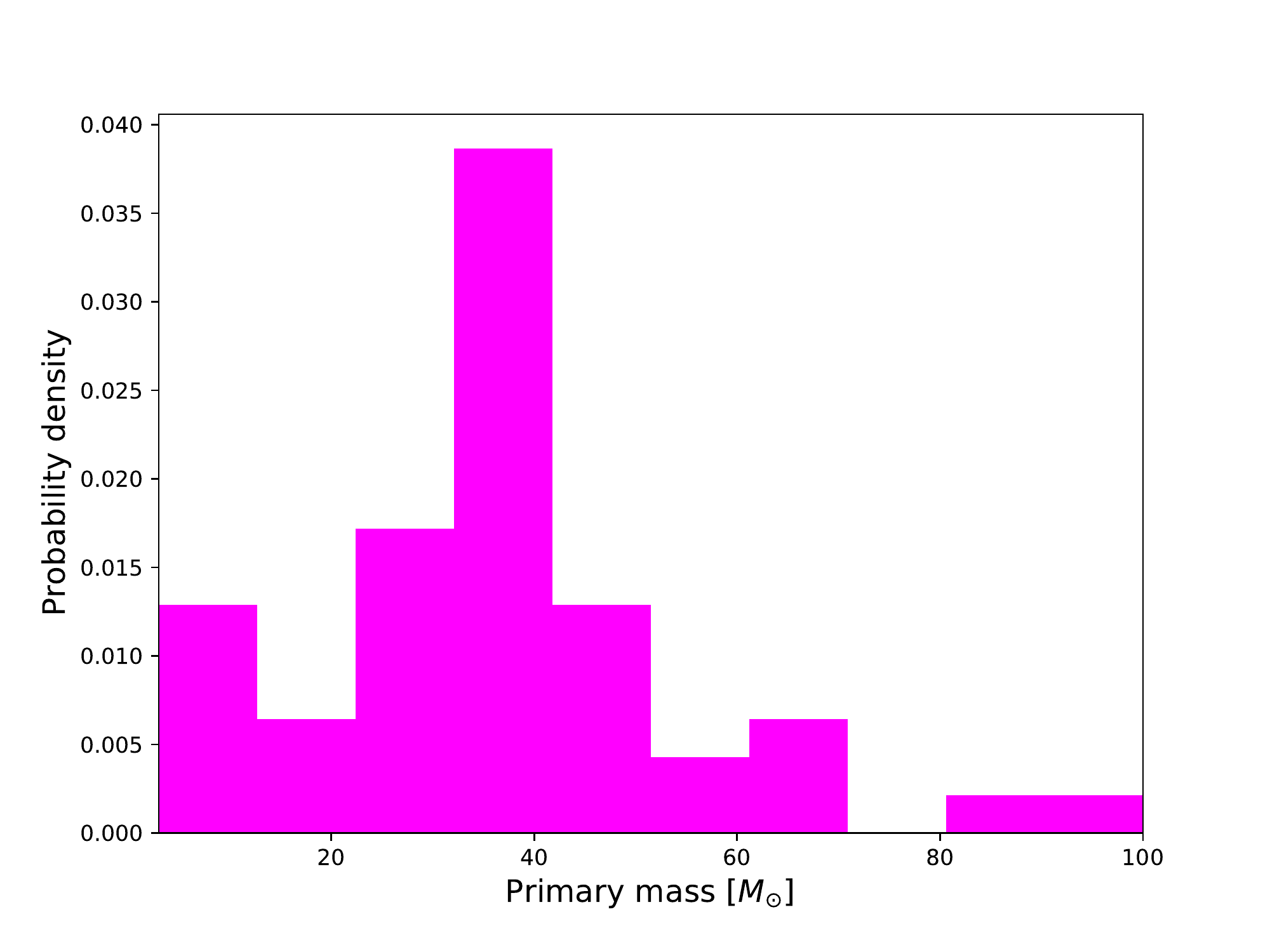}
\includegraphics[width=0.49\textwidth]{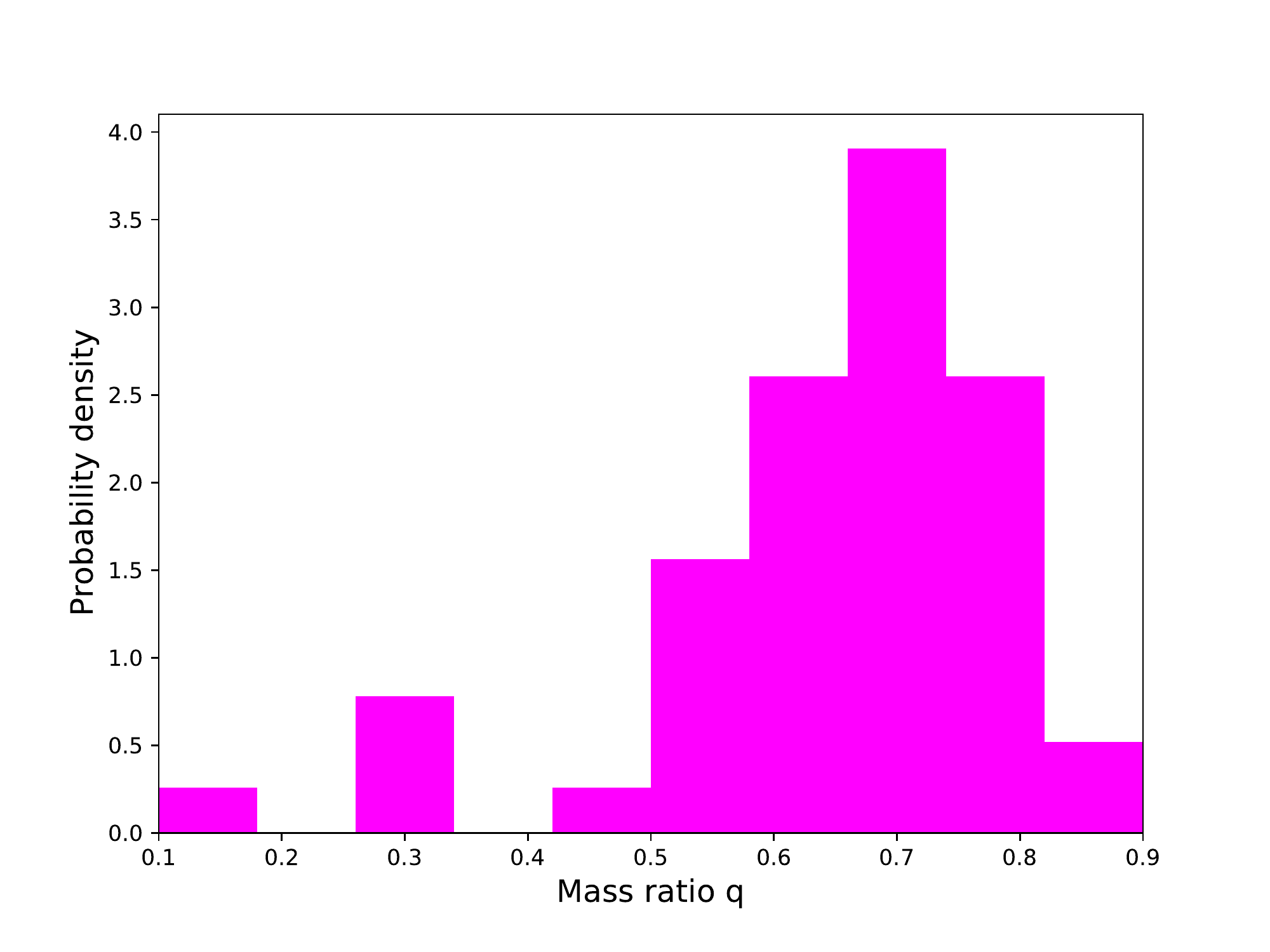}
\caption{The histogram of the primary mass and mass ratio of BBH in GWTC-2. We divide 10 bins uniformly for the primary mass from 3 to 100 $M_{\odot}$ and for the mass ratio from 0.1 to 0.9.}
\label{fig:BHmass}
\end{figure}

The redshift distributions of the BNS and BBH can be sampled from Eq.~(\ref{eq:Rz}). To construct the catalogs we then need to assign the parameters such as mass, sky location, etc. for every merger candidate.
For BNS, we assume a uniform distribution of mass in [1, 2.5] $M_{\odot}$, which is consistent with the assumption for the prediction of the BNS merger rate in GWTC2~\cite{LIGOScientific:2020kqk}. While for BBH, the mass range is very uncertain. In the study of the population properties of GWTC-2~\cite{LIGOScientific:2020kqk}, a series of mass models for BHs such as ``Truncated'', ``Broken Power Law'', ``Power Law + Peak'', and ``Multi Peak''  have been assumed to fit the mass distribution of BBH catalog. In this paper, the underlying mass spectrum for BBH is not our pursuit. To sample the component mass of BBH, we rely on the confirmed BBH events and assume that future detection follows the same mass distribution as the current catalog. This sampling strategy is not sure to be absolutely accurate but should be safe and conservative. We already have around 50 BBH in the catalog, the possible value and the probability distribution of the component mass of BBH can be directly derived from the histogram of GWTC-2. As shown in figure~\ref{fig:BHmass}, we draw the component mass distribution of BBH from the probability density of the primary mass and mass ratio histograms. In principle, the histograms only reflect the mass distribution of the detected GWs. In these confirmed BBH detections,  the component mass is correlated with the redshift (the mass is relatively larger at higher redshift, otherwise it would not be detected). However, to sample the  BBH candidates (not only the detectable events), we should not include the correlation between mass and redshift. At every redshift, we sample the component mass using the same probability distribution in the histograms. It means that we have both small and large-mass binaries at either low or high redshift. The primary mass and mass ratio peak around 30-40 $M_{\odot}$ and 0.7. In our simulation, we have checked that the catalogs of golden dark sirens do not have preference for larger or smaller component mass and mass ration. So the specific strategy of sampling BBH component mass is not very crucial in our study. We only need to make sure that the mass distribution of BBH in our simulation is plausible from current confirmed detections.

Besides the component mass, there are other parameters that should be assigned for the BNS and BBH catalogs.  
In this paper, we adopt the non-spinning waveform with 9 parameters, $\lambda=$\{$\mathcal{M}_c$, $\eta$, $d_L$, $\iota$, $\theta$, $\phi$, $\psi$, $t_c$, $\phi_c$\}. The distributions of chirp mass $\mathcal{M}_c=(m_1m_2)^{3/5}/(m_1+m_2)^{1/5}(1+z)$ and symmetric mass ratio $\eta=m_1m_1/(m_1+m_2)^2$ are translated from the sampled component masses $m_1$ and $m_2$. The luminosity distance $d_L$ is calculated from the sampled redshift by assuming a fiducial cosmological model $\Lambda$CDM with $H_0=67.72~\rm km~s^{-1}~Mpc^{-1}$ and $\Omega_m=0.3104$, corresponding to the mean values obtained from the latest \textit{Planck} TT,TE,EE+lowE+lensing+BAO+Pantheon data combination~\cite{Planck:2018vyg}. We also fix the present CMB temperature $T_{\rm CMB}=2.7255~\rm K$, the sum of neutrino masses $\Sigma_{\nu}m_{\nu}=0.06~\rm eV$, and the effective extra relativistic degrees of freedom $N_{\rm eff}=3.046$, as in the {\it Planck} baseline analysis. The sky localization ($\theta$, $\phi$), inclination angle $\iota$, and polarization $\psi$ are drawn from isotropic distribution. Without loss of generality we set the time and phase at coalescence to be $t_c=\phi_c=0$. In the mid-frequency band, the binaries are largely separated in the inspiral phase and spin-orbit dipole interactions are suppressed. In this paper, since we focus on the luminosity distance and sky localization, we do not include the spin effects of BNS and BBH.
The frequency-domain version of the strain in the inspiral phase reads 
\begin{equation}
\tilde{h}(f)=\sqrt{\frac{5\pi}{24}}\frac{(G \mathcal{M}_c)^{5/6}}{c^{3/2}}\frac{\mathcal{F}}{d_L}(\pi f)^{-7/6}e^{-i \Psi} \,.
\end{equation}
The factor $\mathcal{F}$ is to characterize the detector response, $\mathcal{F}=\sqrt{\frac{(1+\cos^2\iota)^2}{4}F^2_++\cos^2\iota F^2_\times}$. $F_+$ and $F_\times$ are the antenna response functions to the  $+$ and $\times$ polarizations of GW. The single-baseline of AEDGE reorients and moves along the orbit around the Earth, so $\mathcal{F}$ is a function of time and the observed waveform and phase are modulated and Doppler-shifted, yielding important angular information. We follow the same setup for AEDGE as Paper I. The detailed calculations for the antenna response functions can be found there. The specific form of the phase $\Psi$ in the waveform can be found in the appendix of~\cite{Graham:2017lmg}. We expand $\Psi$ to the second PN order as in~\cite{Ruan:2019tje}.
For every merger, the SNR is calculated from $\rho=\sqrt{(h,h)}$. We set a threshold for the detection of GWs, $\rho_{\rm th}>8$. To infer the uncertainty of the parameters in the waveform and the covariance between them, we adopt the fisher information matrix $\Gamma_{ij}=\left(\frac{\partial h}{\partial\lambda_i},\frac{\partial h}{\partial \lambda_j}\right)$. The inner product is defined as
\begin{equation}
(a,b)=4\int_{f_{\rm min}}^{f_{\rm max}}\frac{\tilde{a}^*(f)\tilde{b}(f)+\tilde{a}(f)\tilde{b}^*(f)}{2}\frac{df}{S_n(f)} \,,
\label{eq:product}
\end{equation}
where $S_n(f)$ is the one-sided noise power spectral density (PSD) of detector. We use the sensitivity curve of AEDGE in the resonant modes (see the envelope in figure 1 of~\cite{Ellis:2020lxl}).  The frequency window of AEDGE is from 0.01 to 3 Hz, and the most sensitive region lies in [0.08, 3] Hz. From the fisher matrix, the uncertainty of the parameter is $\Delta\lambda_i=\sqrt{\Gamma^{-1}_{ii}}$. The error of the sky localization is $\Delta\Omega=2\pi|\sin\theta|\sqrt{\Gamma^{-1}_{\theta\theta}\Gamma^{-1}_{\phi\phi}-(\Gamma^{-1}_{\theta\phi})^2}$~\cite{Cutler:1997ta}. 

In the mid-frequency band ($\sim 0.1$ Hz), the time scales of BNS and BBH inspiral phase could be months to years. Considering the limited resource for tracking BNS and BBH in the resonant modes of AEDGE, we set a starting frequency $f_{\rm min}$ in the integral~\ref{eq:product}. For BBH $f_{\rm min}=0.05$ Hz and for BNS  $f_{\rm min}=0.2$ Hz. In both cases, $f_{\rm max}=3$ Hz is the upper limit of the frequency window~\footnote{For BNS or BBH, the freuquency of the Inner-most Stable Circular Orbit at which the inpiral phase ends is usually much larger than 3 Hz.}. The reason for the choices of $[f_{\rm min},f_{\rm max}]$ is that we would like to limit the observation time for a typical BNS or BBH to be around or less than 1 year. For instance, the inspiral time from 0.2 to 3 Hz for a $1.4-1.4$ $M_\odot$ BNS at $z=0.02$ is 1.04 year. In the case of BBH, the inspiral time from 0.05 to 3 Hz for $10-10$ $M_\odot$ and $30-30$ $M_\odot$ binary is 1.4 and 0.22 year, respectively.

The strategy of observing BNS and BBH in the resonant modes of AIs deserves more discussion. There are two operation modes, i.e., broadband and resonant modes~\cite{Graham:2017pmn,AEDGE:2019nxb}. In the broadband modes, AIs operate just like the traditional LIs which can observe the sources in a broad frequency band. While for the resonant modes, the frequency band is very narrow at a specific value. The observing strategy is to first sit at the lower limit of the frequency window, waiting for a source to enter the band. Once a source is discovered it can be tracked for longer by sweeping the detector frequency up to follow the source. In the resonant mode, the resonance frequency can be chosen anywhere in the range between 0.01 Hz to 3 Hz. Switching between the different modes or different resonant frequencies can be done rapidly by simply changing the sequence of laser pulses used, without changing hardware or satellite configuration~\cite{Graham:2017pmn}. The optimal strategy of observing tens and even hundreds of BNS or BBH using the resonant modes in a limited operation time  is a non-trivial issue. In the most pessimistic case, we track only one BNS or BBH at one time and follow the source to a higher frequency until 3 Hz. For 5 years of observational time, we can track at least 5 events. As for tracking more events, the strategy is either building more detectors (since the single-baseline configuration of AIs is much cheaper than LIs) or coming up with a dynamic method that can switch between different frequencies rapidly. The study of the specific strategy is out of the scope of this paper. In this paper, we first estimate the total BNS and BBH that could be observed (pass the detection threshold) by a 5-years data-taking period of AEDGE. Then we calculate the localization volume for each BNS or BBH and choose the golden events which have only one potential host galaxy in the 99\% confidence error volume. We construct the catalogs of golden dark BNS and BBH. The Hubble diagram from these golden dark siren is built accordingly. Finally, we assume the realistic number of the golden dark BNS and BBH that can be tracked by AEDGE from the most pessimistic case (5 events) to the most optimistic case (all golden events), preparing for using them to measure the Hubble constant in Section~\ref{sec:hubble}.

\begin{figure}
\centering
\includegraphics[width=0.45\textwidth]{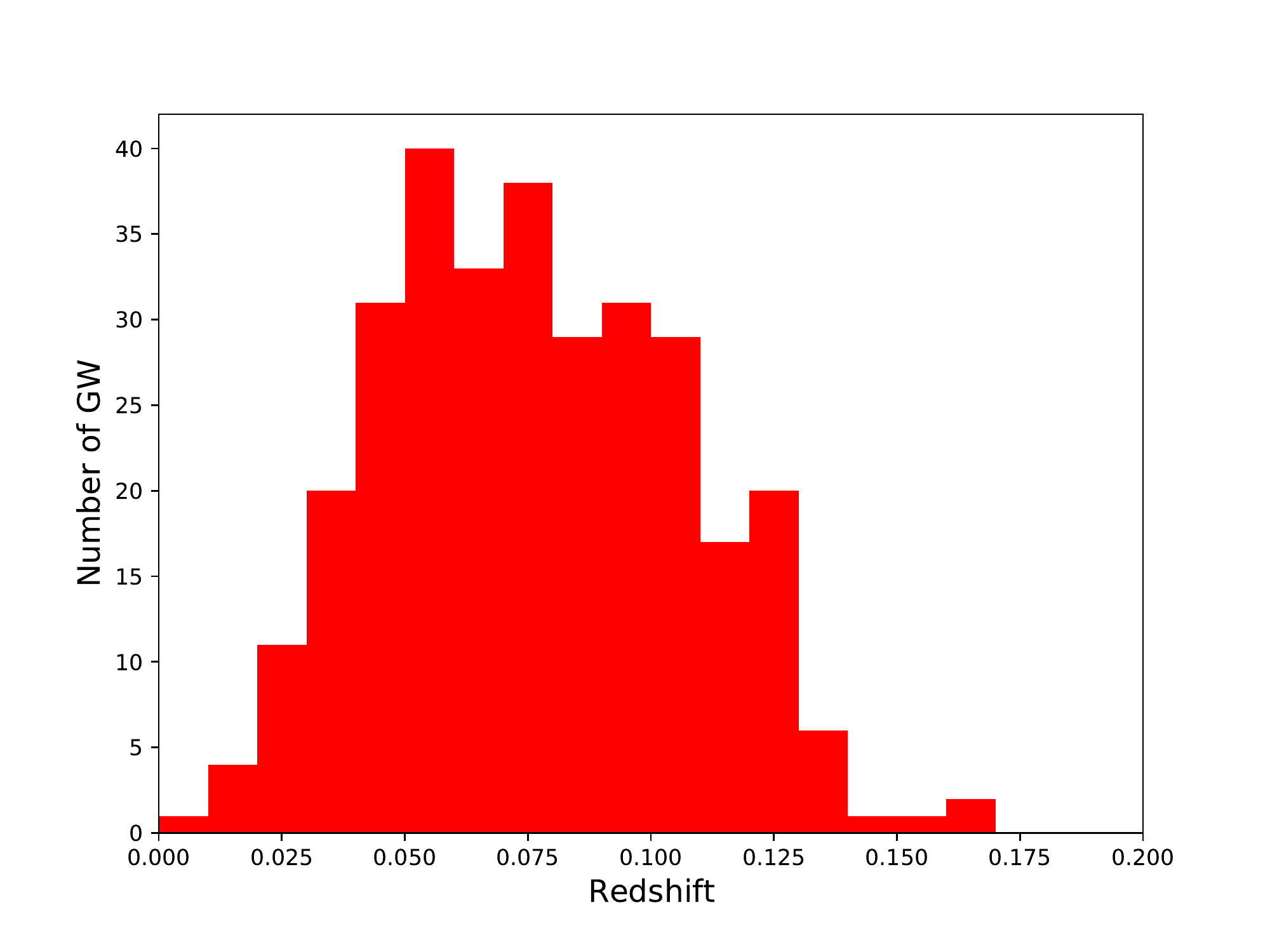}
\includegraphics[width=0.45\textwidth]{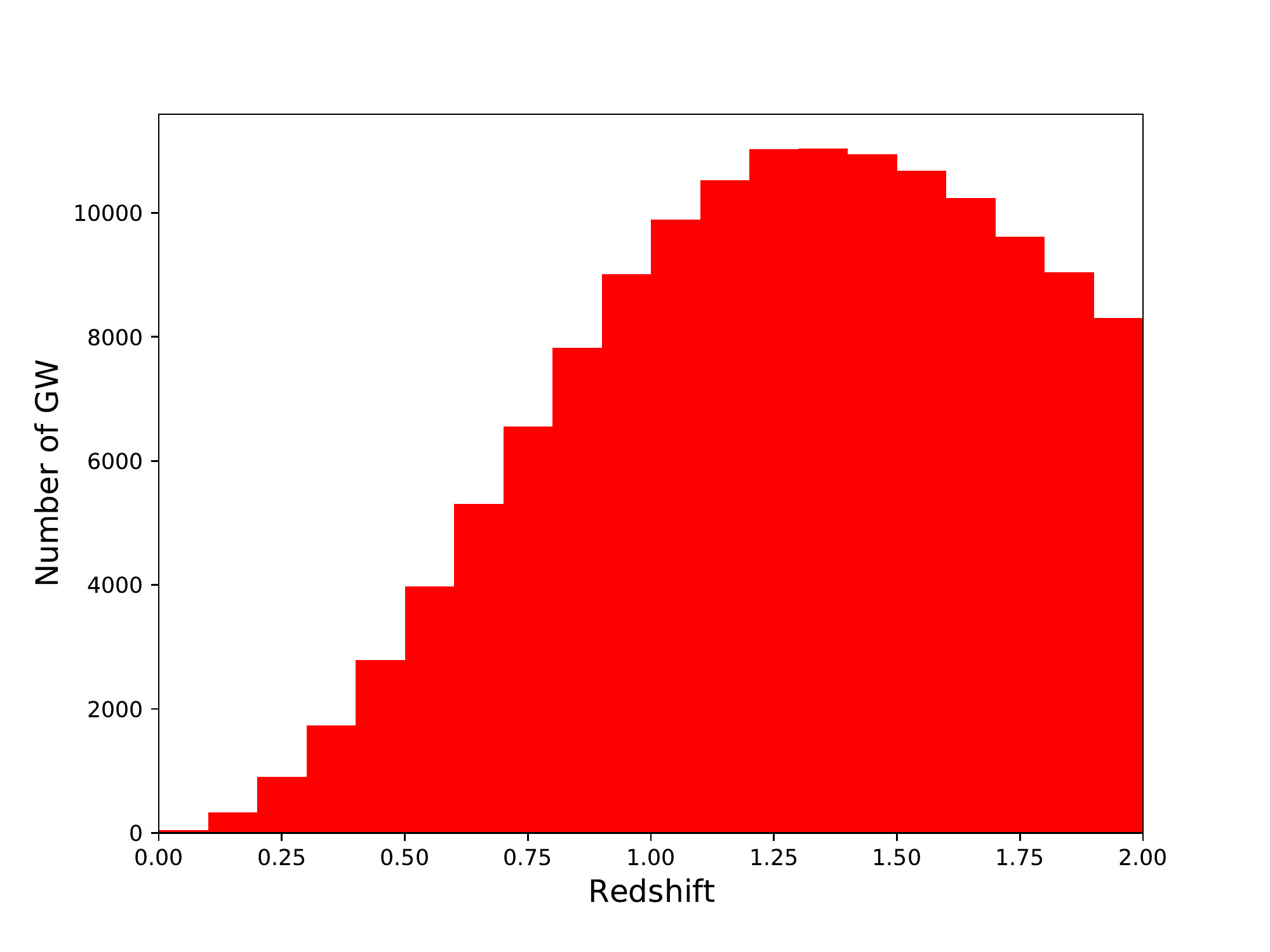}
\caption{The histogram of simulated BNS (left) and BBH (right) that pass the detection threshold (SNR>8) of  AEDGE in 5 years. We set a cutoff at $z=2$ for BBH.}
\label{fig:GWs}
\end{figure}

Figure~\ref{fig:GWs} shows the histogram of simulated BNS and BBH for 5-years of observational time.  There would be around 314 BNS events up to redshift 0.17 that pass the detection threshold of AEDGE. For BBH the number is much larger and about 139818 BBH under redshift 2. The reason we set a cutoff at $z=2$ is as follows. AEDGE can detect  BBH up to redshift greater than 10. But these high-redshift dark sirens are useless (especially in the dark siren case, the source at the large distance is usually poorly localized). Since we only have the low-redshift galaxy catalog and even if we use EM telescope, the high-redshift BBH sources are still not reachable. So we focus on the events below redshift 2 for which we can have the spectroscopic redshift measurement of the host galaxy. 

\begin{figure}
\centering
\includegraphics[width=0.49\textwidth]{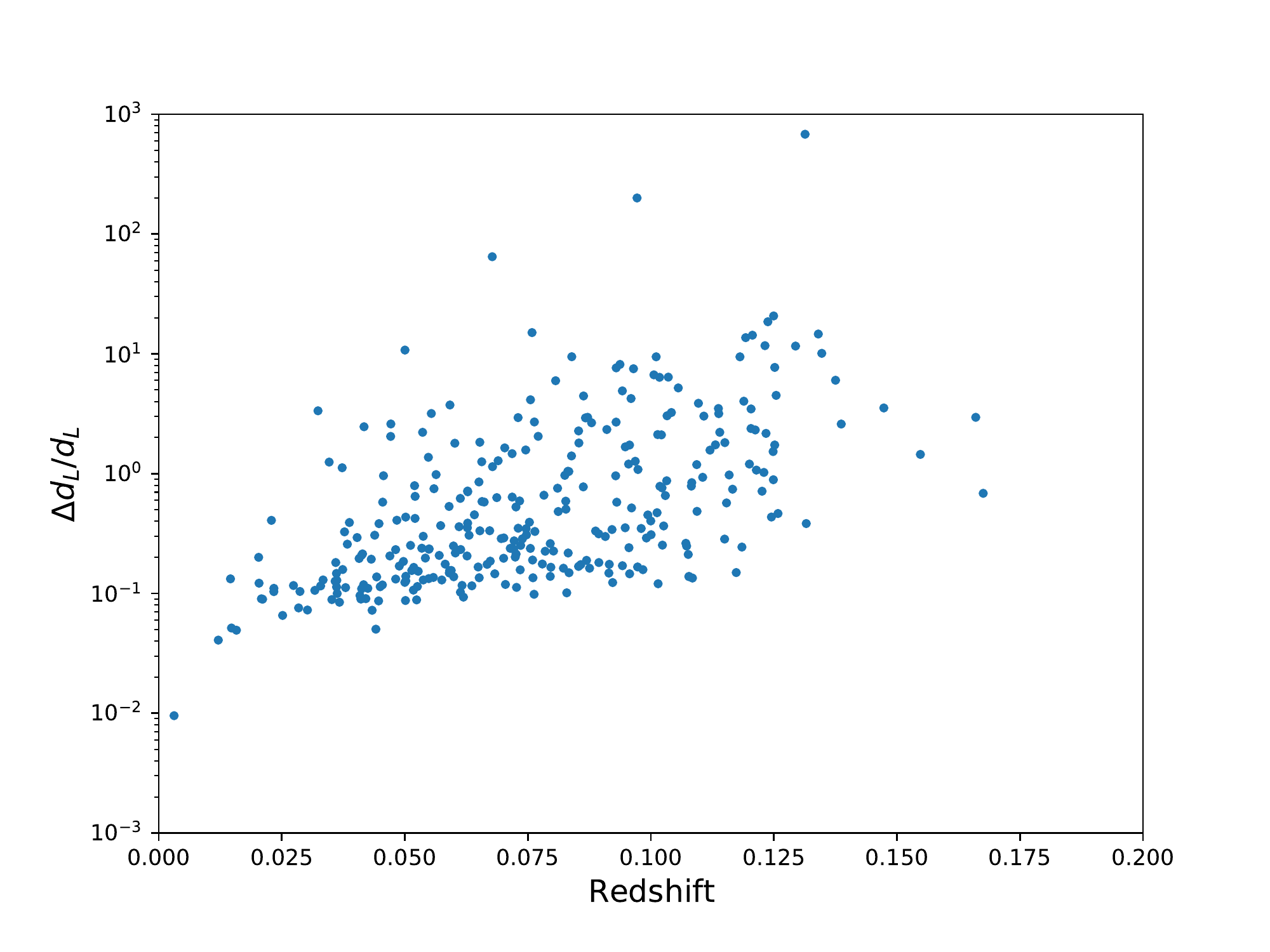}
\includegraphics[width=0.49\textwidth]{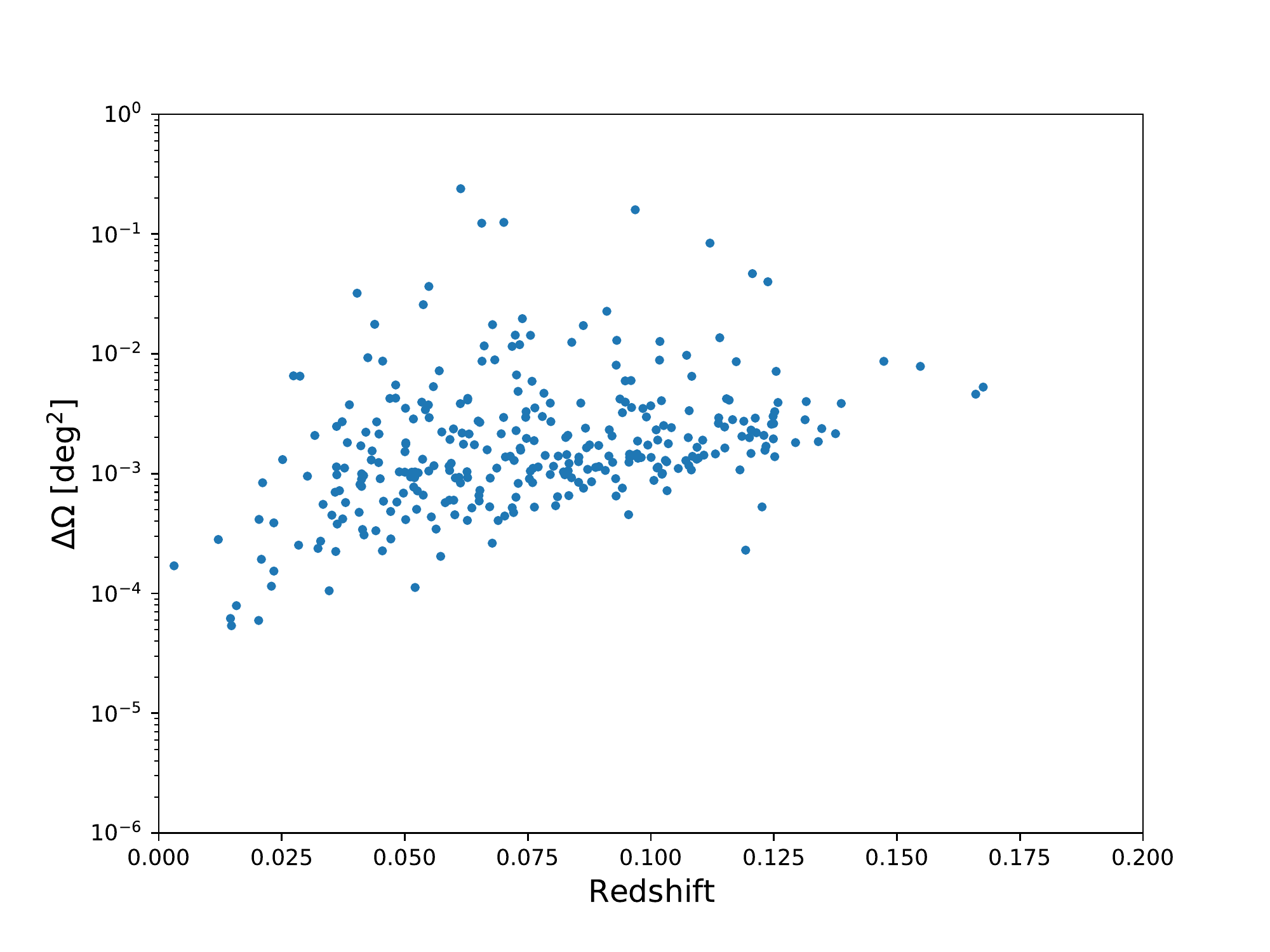} \\
\includegraphics[width=0.49\textwidth]{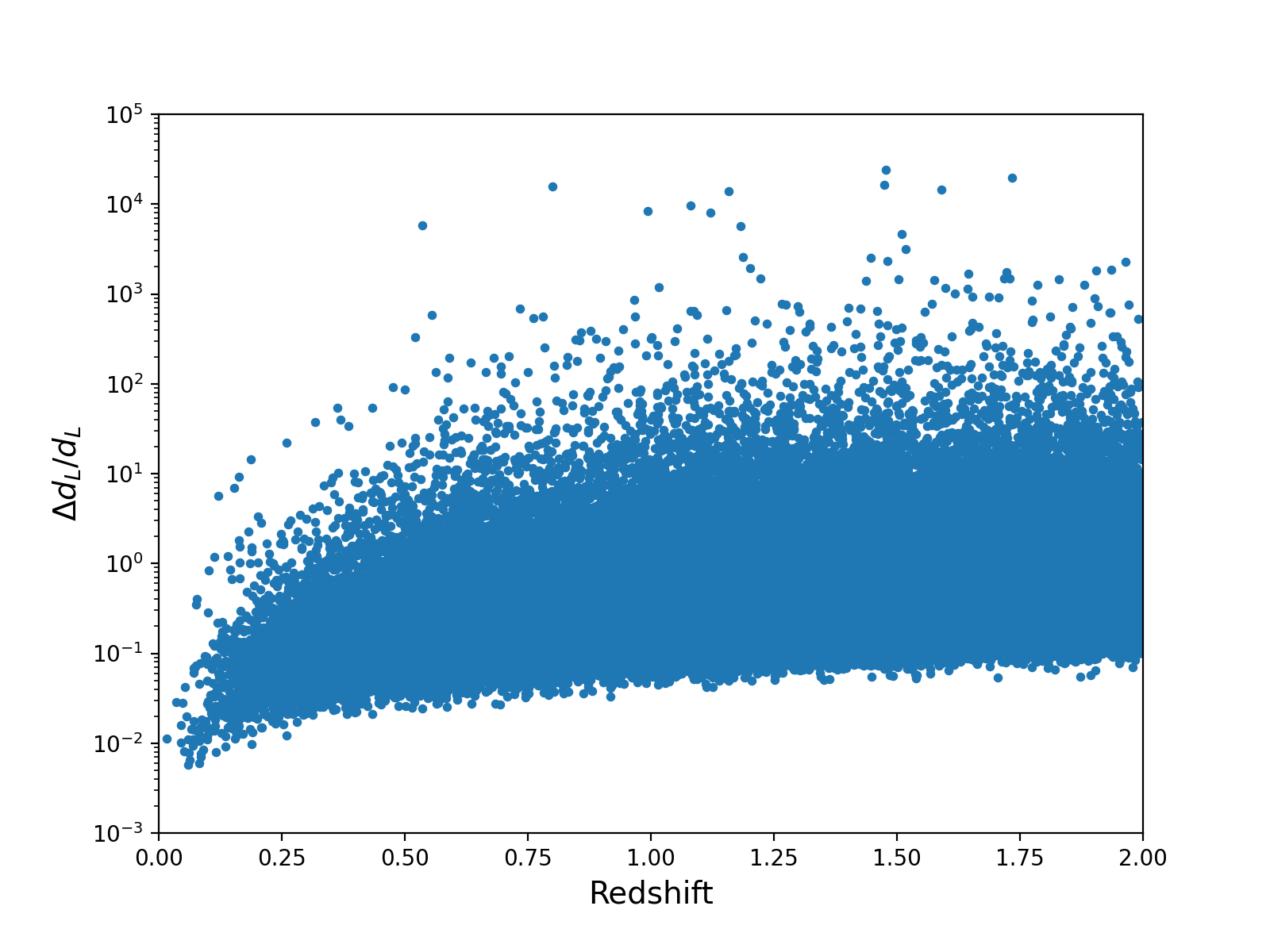}
\includegraphics[width=0.49\textwidth]{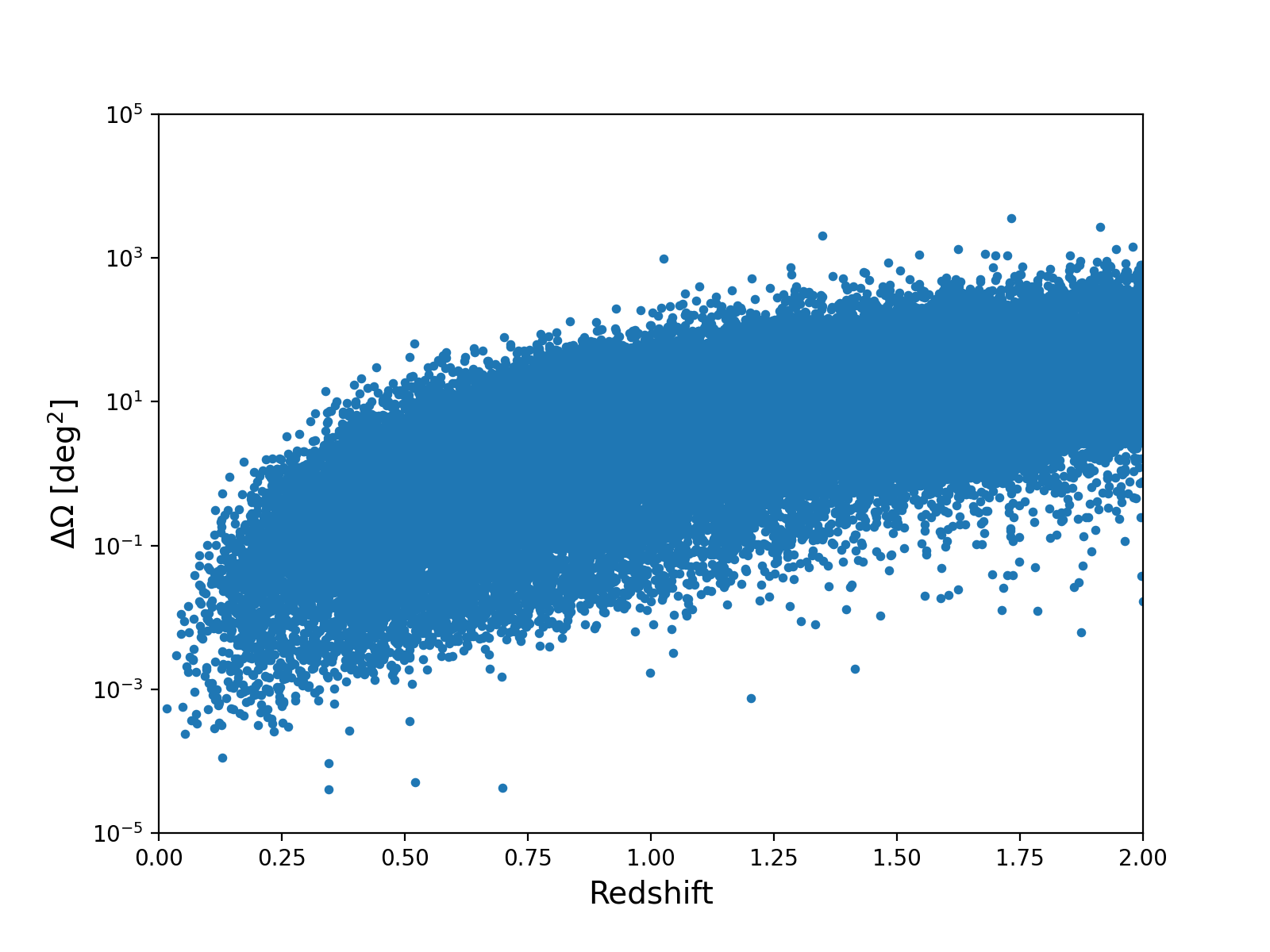}
\caption{The uncertainty of luminosity distance measurement $\Delta d_L/d_L$ and sky localization $\Delta\Omega$ for the simulated BNS (top) and BBH (bottom)  that pass the detection threshold of AEDGE in 5 years. Note the different range of axis between top and bottom panels.}
\label{fig:dL_Omega}
\end{figure}

\begin{figure}
\centering
\includegraphics[width=0.45\textwidth]{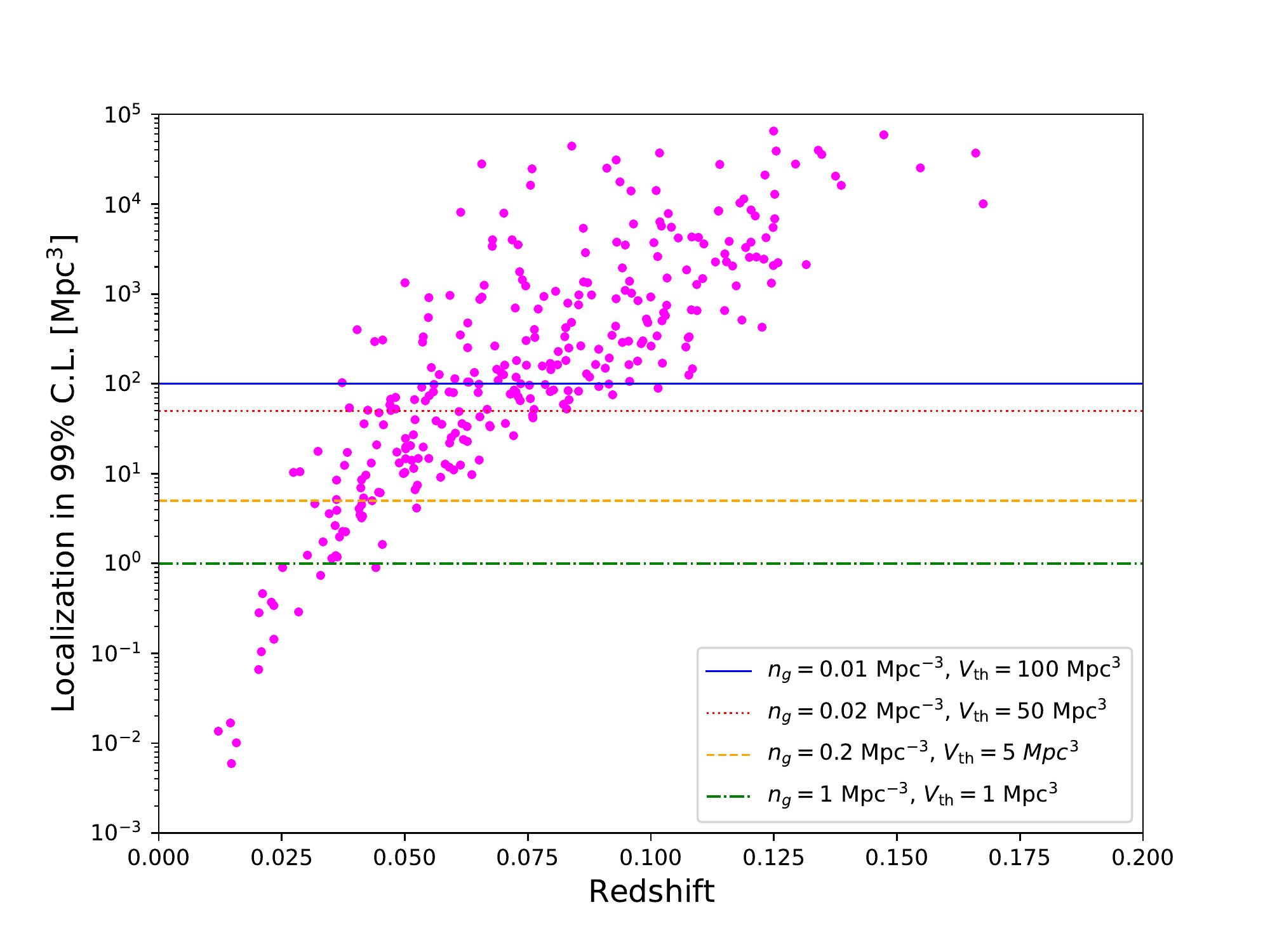}
\includegraphics[width=0.45\textwidth]{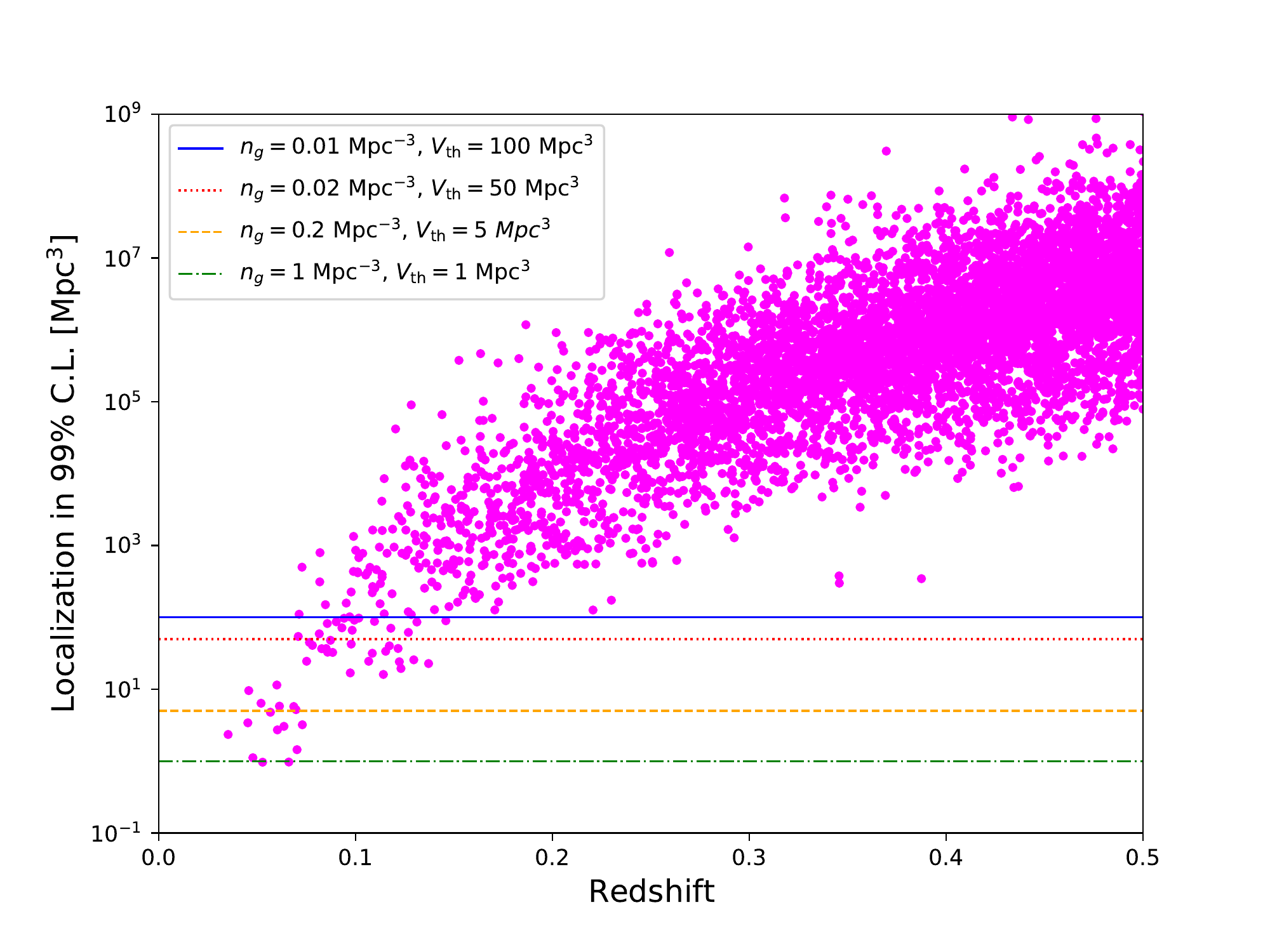}
\caption{The localization volume of the simulated BNS (left) and BBH (right) within 99\% C.L. The lines correspond to the threshold volumes (based on different assumptions) within which there is only one potential host galaxy.}
\label{fig:loc}
\end{figure}

Figure~\ref{fig:dL_Omega} shows the uncertainty of luminosity distance and sky localization for these BNS and BBH which pass the detection threshold of AEDGE in 5 years. We follow the method in~\cite{Yu:2020vyy} to convert $\Delta d_L$ and $\Delta\Omega$ to the 99\% confidence ellipsoid of the localization. We assume the galaxy is uniformly distributed in the comoving volume and the number density $n_g$ is 0.01, 0.02, 0.2, and 1 Mpc$^{-3}$ based on different assumptions, respectively. Then the threshold volume is defined to be $V_{\rm th}=1/n_g$. If the localization volume $V_{\rm loc}<V_{\rm th}$, the host galaxy of this GW can be identified uniquely. The number density of galaxies that host BNS and BBH depends on the specific assumptions and models. In~\cite{Chen:2016tys}, $n_g=0.01~\rm Mpc^{-3}$ by taking the Schechter function parameters in B-band $\phi_*=1.6\times 10^{-2} h^3 {\rm Mpc^{-3}}, \alpha=-1.07, L_*=1.2\times 10^{10} h^{-2} L_{B,\odot}$ and $h=0.7$, integrating down to 0.12 $L_*$ and comprising 86\% of the total luminosity. We start from  $n_g=0.01~\rm Mpc^{-3}$ as the most optimistic case and also set the other three different conservative values. For each $n_g$ we calculate the total number of the golden dark BNS and BBH from the catalogs in figure~\ref{fig:GWs}. However, since AEDGE in the resonant modes can only track a small fraction of the total events, we give our results based on how many golden events can be actually tracked by AEDGE.  Figure~\ref{fig:loc} shows the 99\% error volume of sky localization for these simulated BNS and BBH. The numbers of golden BNS and BBH are summarized in table~\ref{tab:number}. We discard the events with bad constraints of luminosity distance $\Delta d_L/d_L>30\%$, which we think is useless on measuring cosmological parameters.

\begin{table}[]
\centering
\resizebox{\textwidth}{!}{%
\begin{tabular}{|c|c|c|c|c|c|c|}
\hline
\multirow{2}{*}{Number density of galaxies [Mpc$^{-3}$]} & \multicolumn{3}{c|}{$V_{\rm loc}\leq V_{\rm th}$} & \multicolumn{3}{c|}{$V_{\rm loc}\leq V_{\rm th}$ and $\Delta d_L/d_L\leq 30\%$} \\ \cline{2-7} 
                                                         & BNS         & BBH        & Total (BNS+BBH)        & BNS                   & BBH                  & Total (BNS+BBH)                  \\ \hline
0.01                                                     & 135         & 51         & 186                    & 107                   & 49                   & 156                              \\ \hline
0.02                                                     & 96          & 37         & 133                    & 79                    & 35                   & 114                              \\ \hline
0.2                                                      & 36          & 11         & 47                     & 34                    & 11                   & 45                               \\ \hline
1                                                        & 16          & 3          & 19                     & 15                    & 3                    & 18                               \\ \hline
\end{tabular}%
}
\caption{The number of golden GW events under different assumptions of the number density of galaxies.}
\label{tab:number}
\end{table}

Our simulation shows by assuming e.g. $n_g=0.02~\rm Mpc^{-3}$ the number of golden dark sirens of BNS and BBH that pass the detection threshold of  AEDGE are around 79 and 35 in a 5-years data-taking period. The redshifts of these golden dark sirens are up to 0.07 and 0.14 for BNS and BBH, respectively. We would like to mention that the total number of BNS (BBH) that pass the detection threshold of AEDGE below redshift 0.07 (0.14) is around 140 (125). Thus a useful strategy of tracking e.g. the golden dark BNS is focusing on the redshift $z<0.07$ (for GW we use the corresponding luminosity distance). Then the probability of tracking the golden BNS among the total BNS is above 56\% (for golden BBH it is 28\%). Furthermore, by the real-time data analysis in the tracking process, we can predict the quality (also the properties) of the event as soon as possible and improve the successful capture of the golden events.

\begin{figure}
\centering
\includegraphics[width=0.9\textwidth]{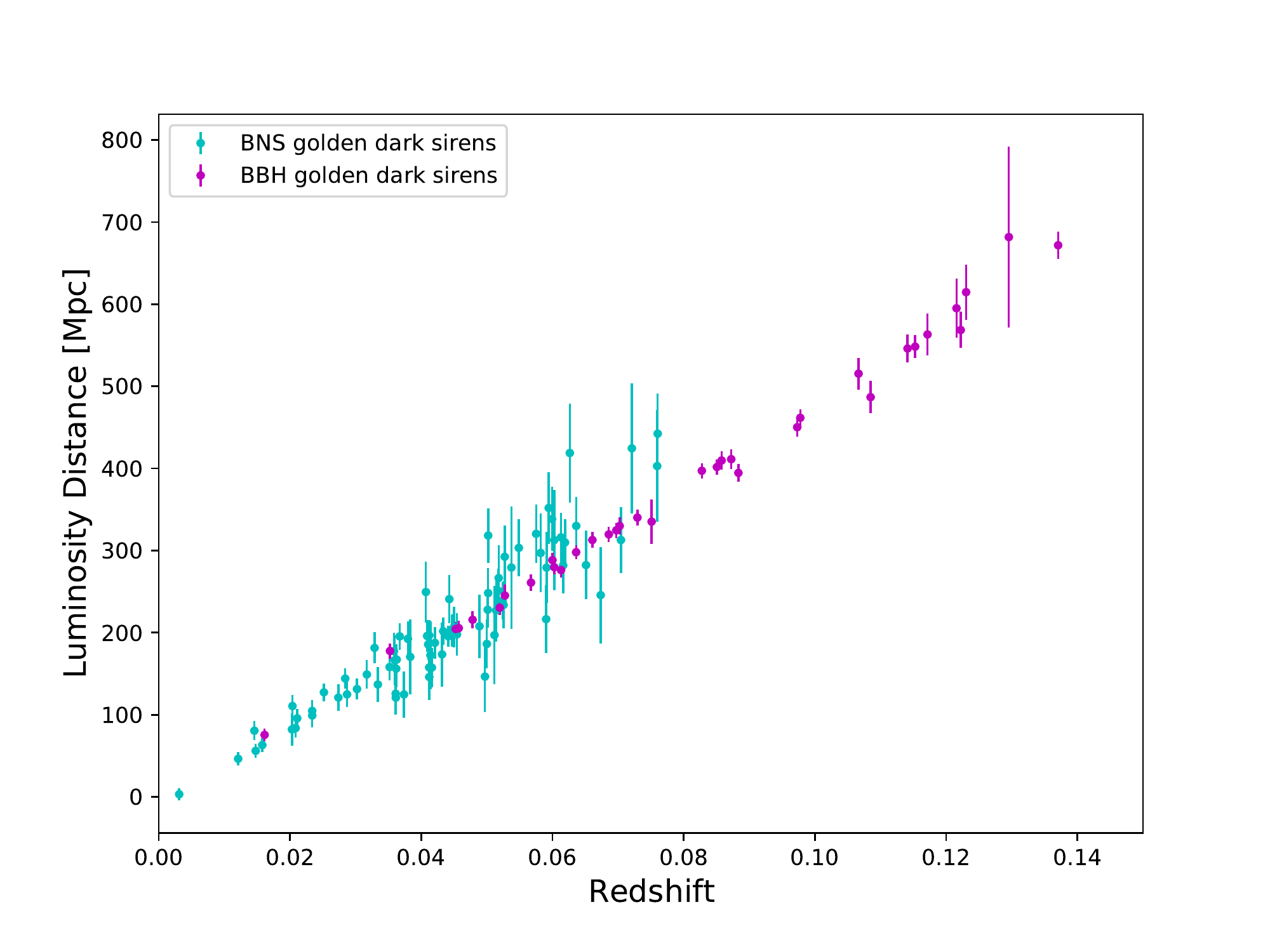}
\caption{The Hubble diagram constructed from one realization of the simulated BNS and BBH golden dark sirens that pass the detection threshold of AEDGE in 5 years. Their host galaxies can be identified unambiguously by assuming $n_g=0.02~\rm Mpc^{-3}$.}
\label{fig:StS_Hubble}
\end{figure}

We finally construct the Hubble diagram which has 114 golden events by assuming $n_g=0.02~\rm Mpc^{-3}$ as shown in figure~\ref{fig:StS_Hubble}. Since the golden dark sirens mainly reside in the low redshift region, we can safely neglect the weak lensing contribution in $\Delta d_L$. While the peculiar velocity of a galaxy is more prominent in small $z$,
we use the fitting formula~\cite{Kocsis:2005vv},
\begin{equation}
\left(\frac{\Delta d_L(z)}{d_L(z)}\right)_{\rm pec}=\left[1+\frac{c(1+z)^2}{H(z)d_L(z)}\right]\frac{\sqrt{\langle v^2\rangle}}{c} \,,
\end{equation}
here we set the peculiar velocity value to be 500 km/s, in agreement with average values observed in galaxy catalogs. We should note that the peculiar velocity is the main error contributor to the redshift measurement of the host galaxy in the local Universe. In this fitting formula, we convert the uncertainty of redshift to that of luminosity distance. The final uncertainty of $d_L$ in the Hubble diagram is the sum of error from the fisher matrix calculation of GW and from the peculiar velocity in quadrature.

\section{The constraints of Hubble constant \label{sec:hubble}}

\begin{figure}
\centering
\includegraphics[width=0.9\textwidth]{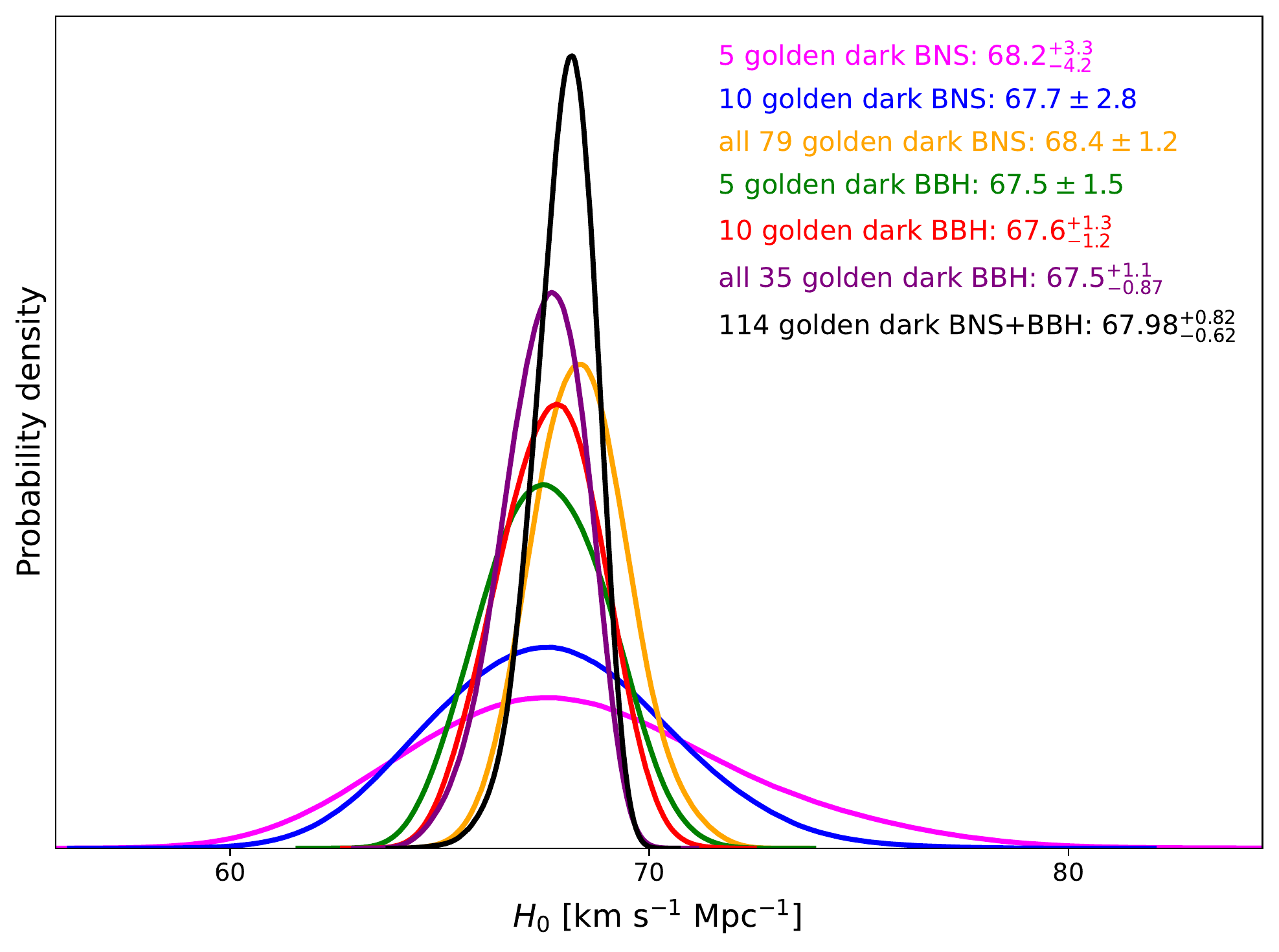}
\caption{The constraints of the Hubble constant by the golden dark sirens of AEDGE.  The numbers are the mean values with 68\% limits of the errors.}
\label{fig:H0}
\end{figure}

The number of golden dark sirens that can be tracked by AEDGE in the resonant modes is still uncertain. In this paper, we consider different cases for the number of golden dark sirens that can be tracked. For comparison, we divide the cases into 5, 10, and all golden events of BNS and BBH. Averagely, the number of golden events that can be tracked by AEDGE in the most pessimistic situation is 1 per year. Thus the numbers 5 and 10 correspond to the 5 and 10-years observational time of AEDGE. However, by optimizing the tracking strategy (especially for BBH whose tracking time is from a few months to 1 year), we can observe more golden events in 5 --10 years. We randomly select these events in the catalogs of golden BNS and BBH. Since the golden events mainly reside in the low redshift region, we focus on constraining the Hubble constant. We assume the $\Lambda$CDM model with two free parameters $H_0,~\Omega_m$. However, at low redshift, $\Omega_m$ is poorly constrained. To get the posteriors of $H_0$, we run Markov-Chain Monte-Carlo (MCMC) by using the package {\sc Cobaya}~\cite{Torrado:2020dgo,2019ascl.soft10019T}. The marginalized statistics of the parameters and the plots are produced by the Python package {\sc GetDist}~\cite{Lewis:2019xzd}. To overcome the bias introduced by the random selection, for each case, we repeat the process (the selection and MCMC) 10 times. We choose the one with the medium relative error among the 10 repetitions as the representative result. The constraints of the Hubble constant are shown in figure~\ref{fig:H0}.

Our results show that with 5, 10, and all 79 golden dark BNS AEDGE can constrain $H_0$ at 5.5\%, 4.1\%, and 1.8\% precision levels (with 68\% confidence level). While with 5, 10, and all 35 golden dark BBH one can constrain $H_0$ at 2.2\%, 1.8\%, and 1.5\% precision levels. Combining all of these golden dark BNS and BBH together, the Hubble constant can be constrained with 1\% precision.  Note when calculating the precision, we use the mean value of upper and lower bounds if the errors are asymmetric. The golden dark BBH is more efficient than BNS in the measurement of the Hubble constant and with only 5-10 golden BBH one can obtain a 2 percent measurement of $H_0$ which is sufficient to arbitrate the current Hubble tension.

\section{Conclusions and Discussions \label{sec:conclusion}}

In this paper, we first construct the catalogs of BNS and BBH dark sirens that pass the detection threshold of the atom interferometer AEDGE in 5 years. To fully take the advantage of AEDEG on the precise localization of the GW sources, we focus on the golden dark sirens which have only one potential host galaxy in the localization volumes. Thus the redshift of these golden dark sirens can be inferred unambiguously. By assuming the average number density of galaxies that are uniformly distributed in the comoving volume, we calculate the number of the golden dark BNS and BBH whose ellipsoid volume is less than the threshold volume such that their host galaxies can be identified unambiguously. We then improve the quality of these golden dark siren by discarding the cases with $\Delta d_L/d_L>30\%$.  Our simulation shows that for $n_g=0.02~\rm Mpc^{-3}$ the number of golden dark BNS and BBH that pass the detection threshold of AEDGE in 5 years are around 79 and 35.  With 5, 10, and all 79 golden dark BNS, AEDGE can constrain $H_0$ at 5.5\%, 4.1\%, and 1.8\% precision levels. Golden dark BBH provide better precision: with 5, 10, and all 35 golden BBH one can constrain $H_0$ at 2.2\%, 1.8\%, and 1.5\% precision levels. It suggests that 5-10 golden dark BBH alone is sufficient to arbitrate the Hubble tension. If we can track all these BNS and BBH, then the Hubble constant can be further constrained with 1\% precision.

The numbers of golden dark sirens summarized in table~\ref{tab:number} are the events that pass the detection threshold of AEDGE in 5 years. In the resonant modes and assuming the pessimistic situation, only a small fraction of them can be tracked. In this paper, we obtain the constraints of Hubble constant based on the specific number of  golden dark siren that AEDGE can eventually track. The optimal strategy of tracking as many GWs as possible is still under debate and we leave this for future research.

We assume four separate number density of galaxies $n_g=$ 0.01, 0.02, 0.2, and 1 Mpc$^{-3}$, and the galaxies are uniformly distributed in the comoving volume. This is somehow simple and ideal comparing to the realistic situation. The number density of the galaxies depends on the mass range of the galaxy that we assume to host BNS and BBH. By assuming a smaller mass of galaxy, the number density is larger.  Furthermore, the clustering and grouping of galaxies make the number of galaxies in the localization volumes more uncertain. However, such properties may significantly facilitate the inference of the redshifts of GWs from the cluster and group of the host galaxy instead of the host galaxy itself~\cite{Yu:2020vyy}. For instance, when applying GW170817 as a dark siren (not use the identification of its host galaxy NGC4993), though there are more than 400 potential host galaxies within its localization region, most of the galaxies' redshifts peak around 0.01 (including its host galaxy NGC4993)~\cite{LIGOScientific:2018gmd}. In addition, an outlier galaxy, which is not part of the group or cluster, can happen to reside in the line of sight of the host galaxy. Then from the inference of the luminosity distance of GW source one can easily exclude that outlier (if we know its redshift or distance) from the potential host galaxies.  In this paper, our calculation shows the total numbers of golden dark sirens are from tens to hundreds. We expect the uncertainty of the number density of galaxies would not affect our main results which are based on a few (5-10) golden dark sirens.

The local merger rates of BNS and BBH we use to estimate the number of dark sirens are the medium values of $\mathcal{R}_{\rm BNS}=320^{+490}_{-240}~\rm Gpc^{-3}~\rm yr^{-1}$ and $\mathcal{R}_{\rm BBH}=23.9^{+14.3}_{-8.6}~\rm Gpc^{-3}~\rm yr^{-1}$. To check the influence of the uncertainty of merger rates on our results, we repeat the construction of the catalogs by adopting the lower and upper bounds of $\mathcal{R}_{\rm BNS}$ and $\mathcal{R}_{\rm BBH}$. The total numbers of BNS ($z\sim 0-0.15$) and BBH ($z\sim 0-2$) are $314^{+533}_{-230}$ and $139819^{+83993}_{-50137}$, respectively. Assuming $n_g=0.02~\rm Mpc^{-3}$, the numbers of golden dark BNS and BBH are $79^{+147}_{-60}$ and $35^{+29}_{-10}$. Again, we emphasize that the uncertainty of the numbers would not affect our main results which are only based on 5-10 golden dark sirens.

We can see in figure~\ref{fig:loc} that the localization volume increases with the redshift. At large distance, the sources' redshifts  can be inferred in a statistic way, which provide a much looser constraint of $H_0$ than the golden events. However, the large number of these events can compensate for this inferiority. In this paper, we consider the resonant modes of AEDGE, which cannot track all these events. We would like to mention that the adoption of our methodology in the broadband modes of AIs is very straightforward, as well as for the LIs GW detectors in the mid-frequency band like DECIGO~\cite{Kawamura:2011zz} and BBO~\cite{Harry:2006fi,Yagi:2011wg}. In this case, all dark sirens in a broad range of redshift can be observed. Then not only the Hubble constant but also the dynamics of dark energy and modified gravity theory can be constrained.


\acknowledgments
We would like to thank Gungwon Kang for helpful discussions.
This work is supported by National Research Foundation of Korea NRF-2021M3F7A1082053.
RGC is supported by the National Natural Science Foundation of China Grants No.11690022, No.11821505, No.11991052, No.11947302 and by the Strategic Priority Research Program of the Chinese Academy of Sciences Grant No.XDB23030100, the Key Research Program of the CAS Grant No.XDPB15, and the Key Research Program of FrontierSciences of CAS. 
SJ is supported by Grant Korea NRF-2019R1C1C1010050. 

\bibliographystyle{JHEP}
\bibliography{ref}

\end{document}